\documentclass[conference]{IEEEtran}
\IEEEoverridecommandlockouts
\usepackage{cite}
\usepackage{amsmath,amssymb,amsfonts}
\usepackage{algorithmicx}
\usepackage{algorithm} 
\usepackage{algpseudocode}
\usepackage{graphicx}
\usepackage{pgfplots}
\usepackage{subcaption}
\usepackage{float}
\usepackage{textcomp}
\usepackage{xcolor}
\usepackage{color}
\usepackage{todonotes}
\usepackage{algpseudocode}
\usepackage{algorithmicx}
\usepackage{algorithm}
\usepackage{algpseudocode}
\usepackage{comment}
\usepackage{subcaption}
\usepackage{multirow}
\usepackage{xcolor}
\usepackage{comment}
\usepackage{marginnote}
\usepackage{makecell}
\usepackage[colorlinks]{hyperref}

\usepackage[font=footnotesize,labelfont=bf]{caption}

\def\BibTeX{{\rm B\kern-.05em{\sc i\kern-.025em b}\kern-.08em
    T\kern-.1667em\lower.7ex\hbox{E}\kern-.125emX}}
\begin{document}

\title{
\fontsize{22}{24}\selectfont
 Water Flow Detection Device Based on Sound Data Analysis and Machine Learning to Detect Water Leakage
}

\author{\IEEEauthorblockN{Hossein Pourmehrani\IEEEauthorrefmark{1},
Reshad Hosseini\IEEEauthorrefmark{2}, and
Hadi Moradi\IEEEauthorrefmark{2}}
\IEEEauthorblockA{\IEEEauthorrefmark{1}
University of Maryland Baltimore County, USA
}
\IEEEauthorblockA{\IEEEauthorrefmark{2}
University of Tehran, Iran
}}

\maketitle

\begin{abstract}
In this paper, we introduce a novel mechanism that uses machine learning techniques to detect water leaks in pipes. The proposed simple and low-cost mechanism is designed that can be easily installed on building pipes with various sizes. The system works based on gathering and amplifying water flow signals using a mechanical sound amplifier. Then sounds are recorded and converted to digital signals in order to be analyzed. After feature extraction and selection, deep neural networks are used to discriminate between with and without leak pipes. The experimental results show that this device can detect at least 100 milliliters per minute (mL/min) of water flow in a pipe so that it can be used as a core of a water leakage detection system.
\end{abstract}

\begin{IEEEkeywords}
\noindent
Acoustic signals, Leakage detection, Intelligent water leak detection, Machine learning, Neural network, Deep learning.
\end{IEEEkeywords}

\section{Introduction}
\label{sec:introduction}
Water leak detection is a crucial task since very limited water resources are available, and any drop of water becomes important \cite{b1}. Furthermore, water leaks can damage buildings, structures, and roads. In other words, corrosion and soil movement are expected consequences of leakage, which can cause significant damage to the surrounding infrastructure and buildings, inconvenience customers, and result in financial losses \cite{b1}.

Unfortunately, a majority of water leaks happen through water distribution networks. For instance, in 2020, about 8.4 billion cubic meters of purified drinking water were supplied in Iran, of which 2.3 billion cubic meters were wasted. Of this total waste, 60\% was related to leakage in the transmission line, tanks, valves, and pipes \cite{b2}. Moreover, approximately one-third of global water utilities are estimated to experience a loss of around 40\% due to leakage \cite{b3}.

Generally, there are three types of leaks: inherent, visible, and non-visible \cite{b4}. Inherent leaks are small and cannot be seen or detected, such as those caused by initial corrosion. Visible leaks are easily noticeable as they run on the ground and can be detected by the public. Non-visible leaks cannot be seen but can be identified by their characteristics using specific devices.

There are different devices designed to detect water leaks which mainly work based on the fact that water leakage means water flows from the main valve up to leak points. Systems introduced in \cite{b3} and \cite{b5} use a Flow Liquid Meter (FLM) to detect water flow to identify leakage. One drawback of FLM-based systems is the need to install the system on the water distribution system. The Ultrasonic Flow Meter (UFM) has been designed to solve the previous problem \cite{b6}. In other words, it is a non-invasive system compared to the FLM systems. Unfortunately, UFM-based systems are expensive and not affordable.

\begin{figure}[t]
  \centering
  \includegraphics[width=\columnwidth]{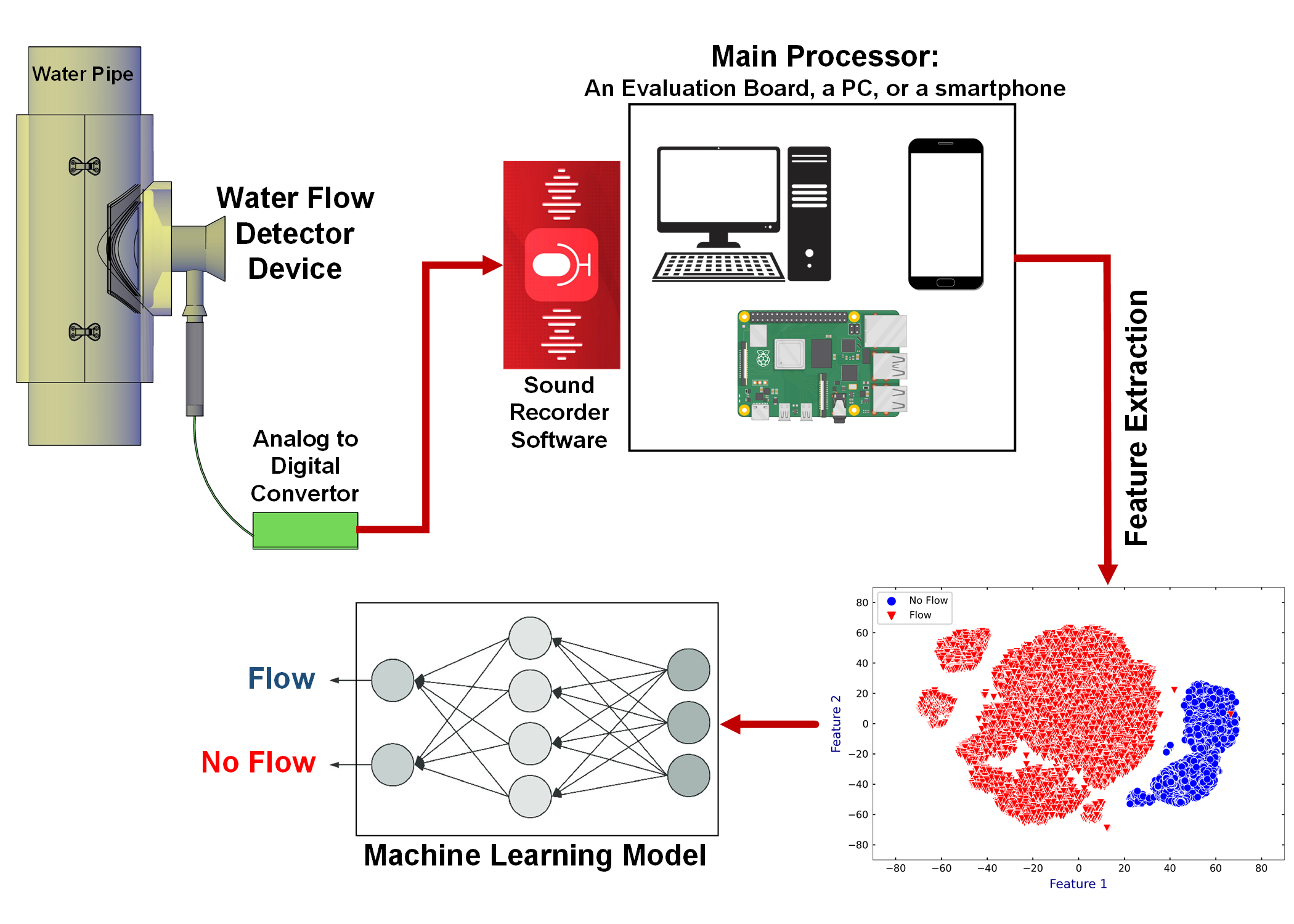}
   \caption{An overview of the proposed water leakage system detection.}\label{fig:overview1}
\end{figure}

Another approach is based on acoustic signals generated at leakage points. Some studies use Hydrophone \cite{b7,b8} to record sound from leakage points inside pipes, while other studies use Geophone \cite{b4, b9} for leakage sound detection. Other standard sensors are vibration \cite{b10,b11}, and acceleration sensors \cite{b12} to record leakage-generated acoustic signals. The water flow at the point of leakage produces sound signals along the pipe, which eventually leads to vibrations in the wall of the pipe. Therefore, vibration and accelerometer sensors are used in this area \cite{b10}. Generally, a vibration sensor is a kind of accelerometer, and in the mentioned studies, a piezoelectric transducer has been used to record vibration signals. Although these systems are precise, multi-sensors need to be installed to detect and localize leakage, and they are highly temperature sensitive.

The main disadvantages of the mentioned technologies are in the price and/or being invasive to the water pipe systems. In contrast, deep learning models can compensate for some of these drawbacks as they have garnered significant attention in recent years due to their exceptional accuracy and ability to solve complex problems that require substantial computational resources~\cite{b13}. These models have found extensive applications across diverse fields, including image processing, sound analysis, and fault detection. Thus, in this study, as shown in Fig.~\ref{fig:overview1}, we have proposed a non-intrusive, simple, and cost-effective design based on the acoustic features of leakage water flows and developing machine learning approaches to classify water leak cases from non-water leak cases with 97.7\% Accuracy.

\begin{figure}[h]
\centerline{\includegraphics[width=\columnwidth]{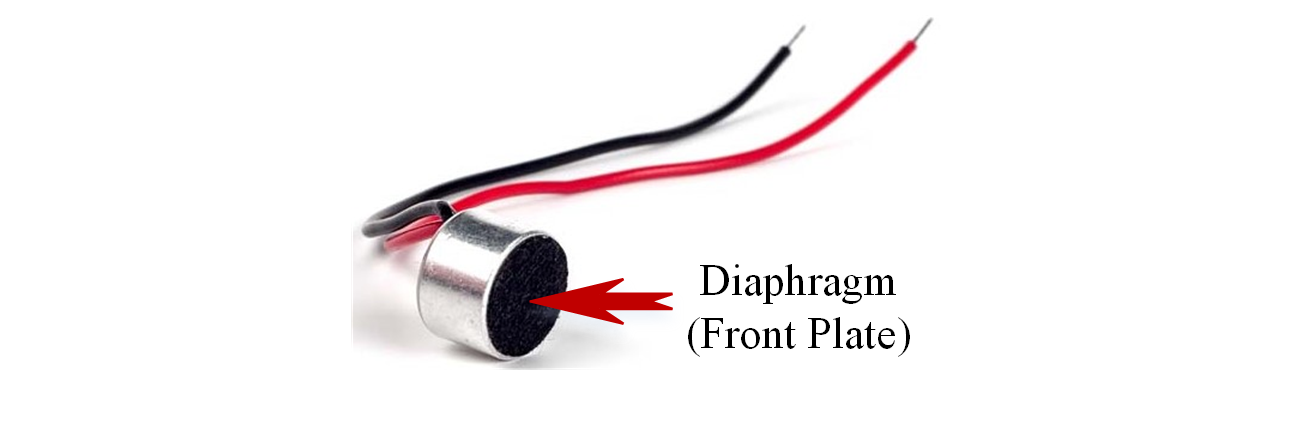}}
\caption{A sample of an acoustic sensor used in the proposed water flow detection system.}
\label{fig:fig1}
\end{figure}

\section{Materials And Methods}
\subsection{Proposed sound water flow meter}
The proposed device works based on the fact that when water flows, it creates sound in pipes carrying water. When there is no actual use of water through a building, which we call the idle situation, there should be no water flow sounds. Thus, if there are such sounds in idle situations, such as midnights, then it should be due to water leakage. Some studies \cite{b4} reported 5 KHz sound frequency that can be generated by water flow. This frequency range can vary based on the water pipe system structure, pipe thickness, and materials. On the other hand, an electret microphone is a sensor that typically records sound from 20 Hz to 20 KHz. Therefore, it can be used to record water flow sounds. Using a microphone has some advantages, including the capability of being installed outside a pipe without any changes in the pipe system and being very cheap compared to other sensors in this field. In Fig.~\ref{fig:fig1}, a sample proposed sound sensor is shown.

\begin{table}[h]
\scriptsize
\centering
\caption{Comparison of the Performance of Each Extraction Feature Method.}
\label{tab:tab1}
\resizebox{250pt}{!}{
\begin{tabular}{ccc}
\hline
\textbf{Feature Extraction Method}          & \textbf{SVM}     & \textbf{Random Forest} \\ \hline
Short Time Fourier Transform (STFT)         & 67.16\%          & 56.06\%                \\
Discrete Wavelet Transform Packet (DWT)     & 64.99\%          & 51.63\%                \\
Mel-frequency cepstral coefficients (MFCCs) & 80.45\%          & 79.50\%                \\
\textbf{FBANK}                              & \textbf{85.50\%} & \textbf{82.10\%}       \\ \hline
\end{tabular}}
\end{table}

\begin{figure}[!t]
\centerline{\includegraphics[width=\columnwidth]{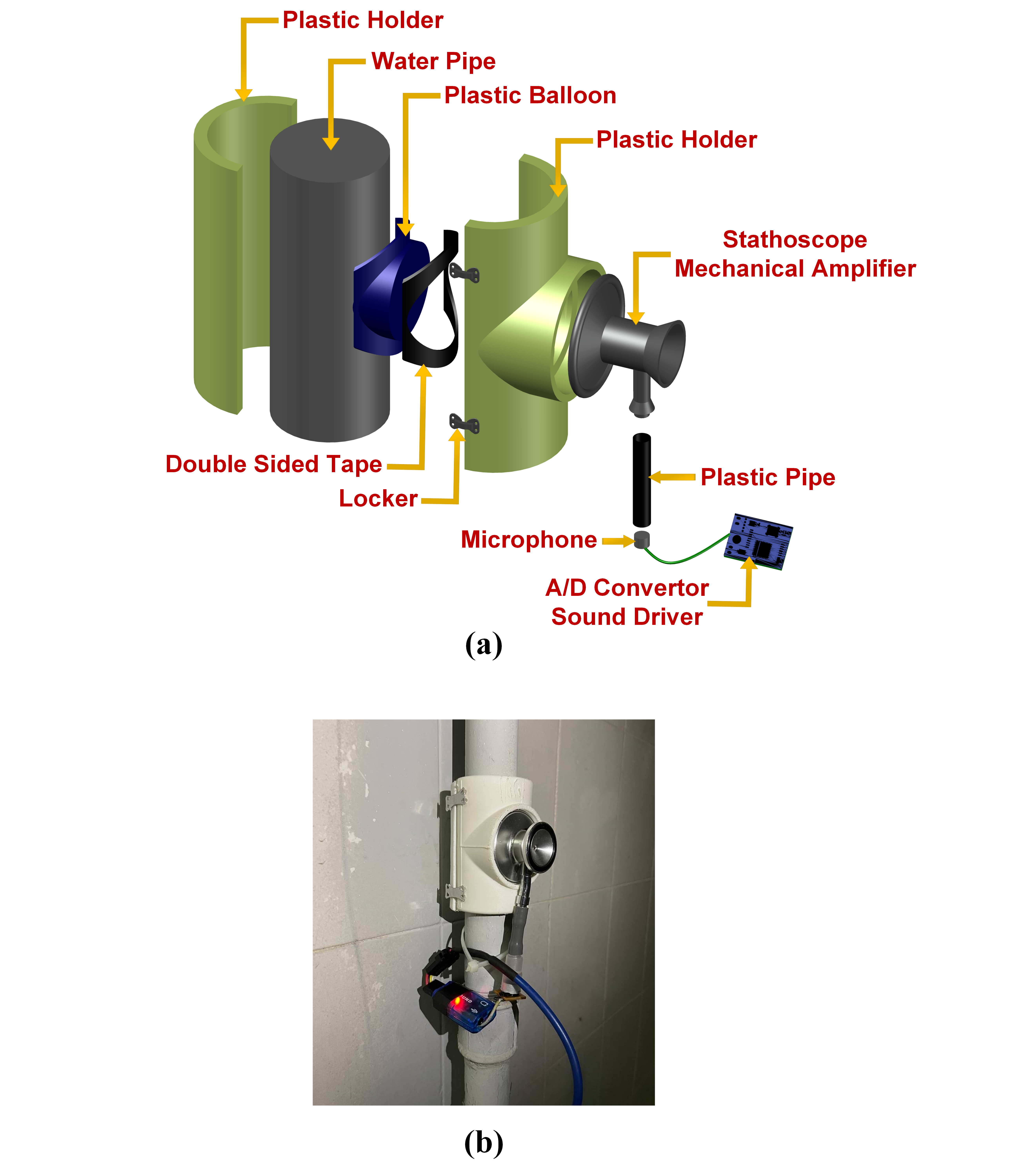}}
\caption{Proposed system (a) CAD model; (b) Prototype.}
\label{fig:fig2}
\end{figure}

Using a piezoelectric or membrane-based microphone would have several challenges. First, direct contact of a microphone with the outer surface of a pipe may change the microphone recording capabilities. In addition, its diaphragm is sensitive to pollution, temperature changes, and dust. Therefore, connecting a microphone directly to the outer surface is not a good option. On the other hand, the sounds generated by water flow especially flows generated due to leakage, are normally weak and need to be amplified. Thus, we have used a cone-shaped structure inspired by stethoscope contact points to human bodies \cite{b14}. Since a stethoscope thin membrane is sensitive to being directly in touch with a hard surface, we have used a plastic balloon filled with a conducting material, such as ultrasound gel, between the membrane and the pipe to prevent membrane damage. This soft material not only takes care of the membrane but also results show that it can transmit acoustic signals from the pipe to the mechanical amplifier very well. Eventually, the entire set can be connected to the outside of the pipe by a plastic pipe support bracket shown in Fig.~\ref{fig:fig2}(a).

In summary, we achieve several goals by implementing such a device. Firstly, we prevent direct contact of the microphone with the pipe and its malfunction. Secondly, using a mechanical amplifier will increase the quality of the audio signals, which helps the microphone detect water flow. Finally,
it is a noninvasive mechanism that can be easily installed on the outside of the pipe without any water pipe changes.

\subsection{Hardware Design}
Fig.~\ref{fig:fig2}(a) illustrates the CAD model of the main parts of the acoustic water flow detection system. All components are connected to the outer surface of the pipe by a plastic pipe support bracket, which consists of two separate parts. It is feasible to change the size of the bracket according to the diameter of pipes between 20 mm and 60 mm. The components of this system consist of various parts as shown in Fig.~\ref{fig:fig2}. The internal part comprises a plastic balloon connected to the pipe by double-sided adhesive tape. This soft balloon connects the membrane of the mechanical amplifier to the pipe to transmit sound signals. The external parts include the rest of the components; the mechanical amplifier is the most important, connected to the soft plastic balloon through a hole that hosts the mechanical amplifier through a plastic tube. Finally, the microphone is connected to an analog-to-digital converter or an audio driver card, whose output is digital audio signals for recording and processing. This device can connect to any processing unit via a USB connection. The prototype of the system in the real application is shown in Fig.~\ref{fig:fig2}(b).

\begin{figure}[!t]
\centerline{\includegraphics[width=\columnwidth]{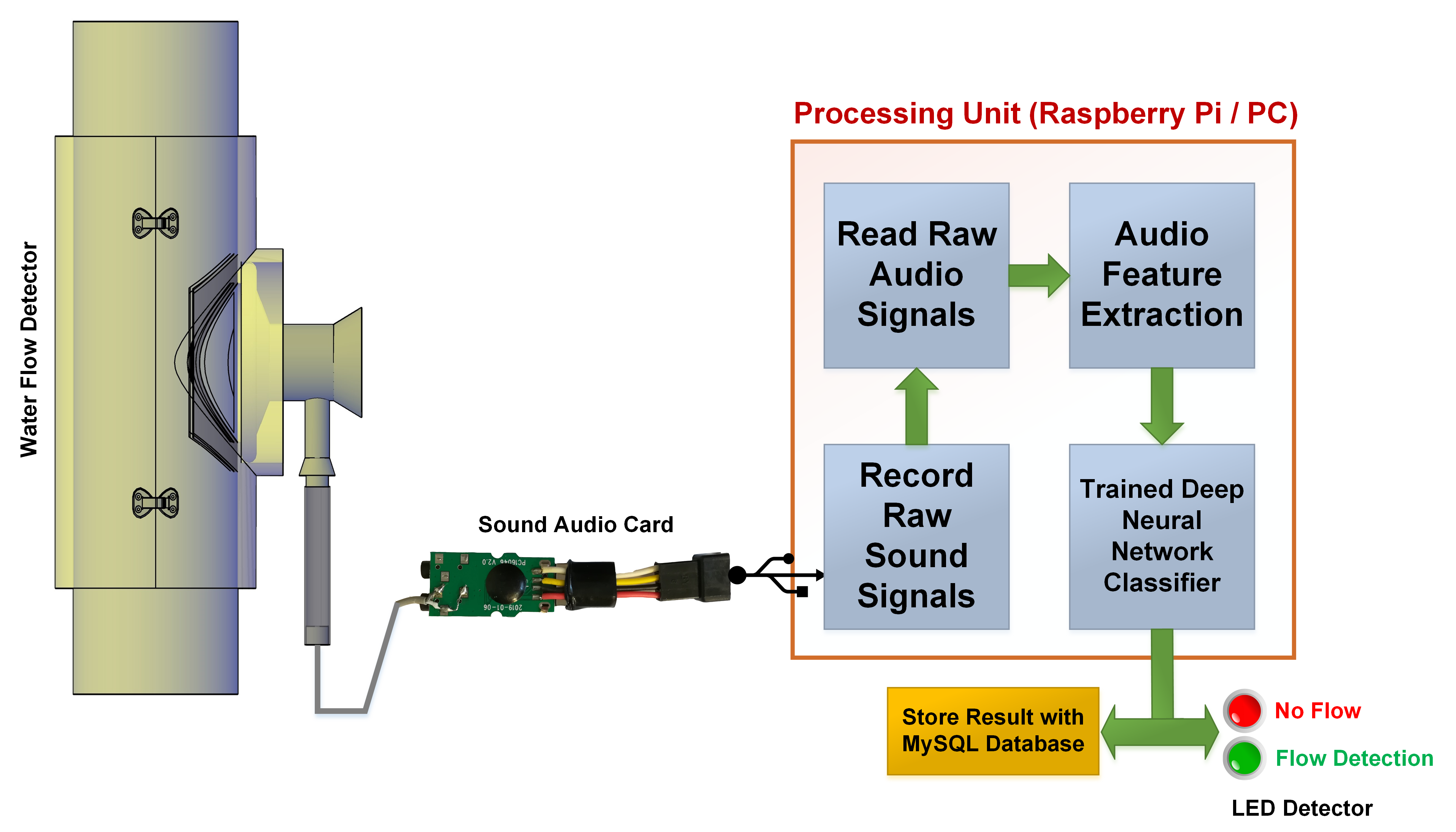}}
\caption{Processing block diagram to detect water flow.}
\label{fig:fig3}
\end{figure}

Fig.~\ref{fig:fig3} illustrates the processing block diagram of the proposed system. The water flow detection system is attached to the outside of the pipe. Then the received water flow sound signals are converted into digital signals via the audio driver. The raw digital audio signals are sent to a processing system, which can be a general-purpose computer, SBCs like Raspberry Pi, or any customized designed computer. We have used a Raspberry Pi board to process data.

\vspace{1.0em}
\subsection{Data Gathering and Feature Extraction}

As mentioned earlier, audio signals are converted into digital form. Then, to train an efficient classifier to distinguish between water leak and no water leak cases, we had to collect adequate data and extract reliable features to train different classifiers. Thus, data were collected using different water flows in a controlled setup rather than testing water leakage in actual setups. The idea was to train classifiers to detect water flow so they can be used to determine water flow in case of leakage versus non-flow cases.

We used the same system described in Fig.~\ref{fig:fig2}(b) to collect data. Various water flows (i.e. 50, 100, 250, 500, 1000, and 2000 mL/min) were produced by a faucet in the water supply system, and for each flow, we recorded produced sound. This data was collected on one water pipe system, but the system was tested for over 10 days at different times and dates. Audio signals were sampled 44.1 kHz rate. We collected 30 to 40 minutes of data for each water flow class.

After collecting data, features were extracted for analysis and for training machine learning algorithms. Various methods such as Short Time Fourier Transform (STFT), Discrete Wavelet Transform Packet(DWT) \cite{b15}, Mel-frequency cepstral coefficients (MFCCs) \cite{b16}, and FBANK \cite{b17} have been used to extract features. In order to evaluate the best method of feature extraction, the features were extracted in each method and divided into training and testing groups. Then, data were classified into six classes of water flows, i.e., no flow case, 100, 250, 500, 1000, and 2000 mL/min, by using the Support Vector Machine (SVM) \cite{b19} and Random Forest \cite{b20} algorithms. Table.~\ref{tab:tab1} shows the results of classifications for each method. The FBANK method performed better despite being very close to the MFCC method, which indicates that the features extracted with the FBANK were more meaningful than others. In addition, the visualization of data, shown in Fig.~\ref{fig:fig5}, represents that the FBANK features performed well and were selected as the final set of features. In the FBANK method, the extracted features have a dimension of 134, and each feature vector represents the features extracted from a one-second frame of recorded data.
\vspace{0.8em}
\begin{table}[h]
\scriptsize
\centering
\caption{Classification Accuracy of Flows of 50, 100, and 250 mL/min Compared to Zero Flow Using SVM Algorithm.}
\label{tab:tab2}
\resizebox{250pt}{!}{
\begin{tabular}{ccc}
\hline
\textbf{Number}          & \textbf{Flows Classification}     & \textbf{SVM Result} \\ \hline
1         & 50 mL/min vs No Flow          & 38.7\%                \\
2     & 100 mL/min vs No Flow          & 91.6\%                \\
3 & 250 mL/min vs No Flow          & 94.8\%             
\\ \hline
\end{tabular}}
\end{table}

\section{Result And Discussion}

\subsection{Data Classification}

\begin{figure}[!t]
\centerline{\includegraphics[width=\columnwidth]{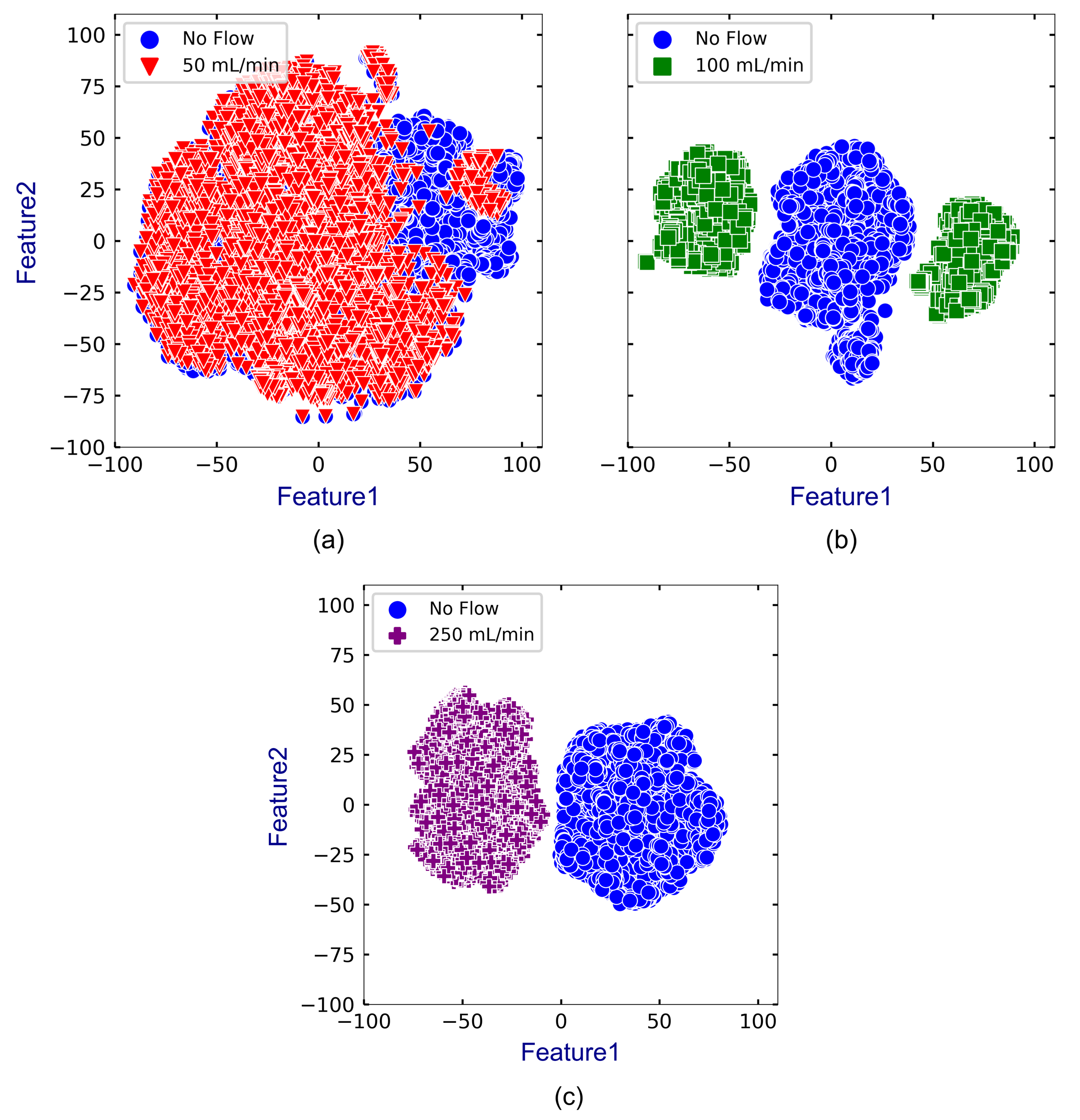}}
\caption{Visualization of different flows in two dimensions using the t-SNE algorithm (a) comparing the flow of 50 mL/min with zero flow; (b) comparing the flow of 100 mL/min with zero flow; (c) comparing the flow of 250 mL/min with zero flow.}
\label{fig:fig4}
\end{figure}

Different classifiers were trained to select the best classifier that can classify water flows in the first three classes of water flows, i.e., 50, 100, and 250 mL/min, with respect to no flow case. In Fig.~\ref{fig:fig4}, mentioned flows are shown in comparison with zero or absence flow by the t-distributed stochastic neighbor embedding (t-SNE) \cite{b18} algorithm in two dimensions. In Fig.~\ref{fig:fig4}(a), zero and 50 mL/min water flow data are displayed that are completely overlapping and cannot be recognized by the system. In contrast, Fig.~\ref{fig4}(b) and (c) illustrate the difference between 100 and 250 mL/min flows compared to the zero flow. Since it is possible that 50 mL/min and zero flow can be separated in a higher dimension, we used the Support Vector Machine (SVM) algorithm \cite{b19} to classify these data. The results are reported in Table.~\ref{tab:tab2}, which shows that it is impossible to classify 50 mL/min and zero flow, while it can be done for 100 and 250 mL/min. In other words, 50 mL/min flow compared to zero flow is not meaningful. Thus, our system can detect water flows or leakage of more than 100 mL/min.

\vspace{1.0em}
\subsection{Data Visualization and Classification}
\begin{figure}[!t]
\centerline{\includegraphics[width=\columnwidth]{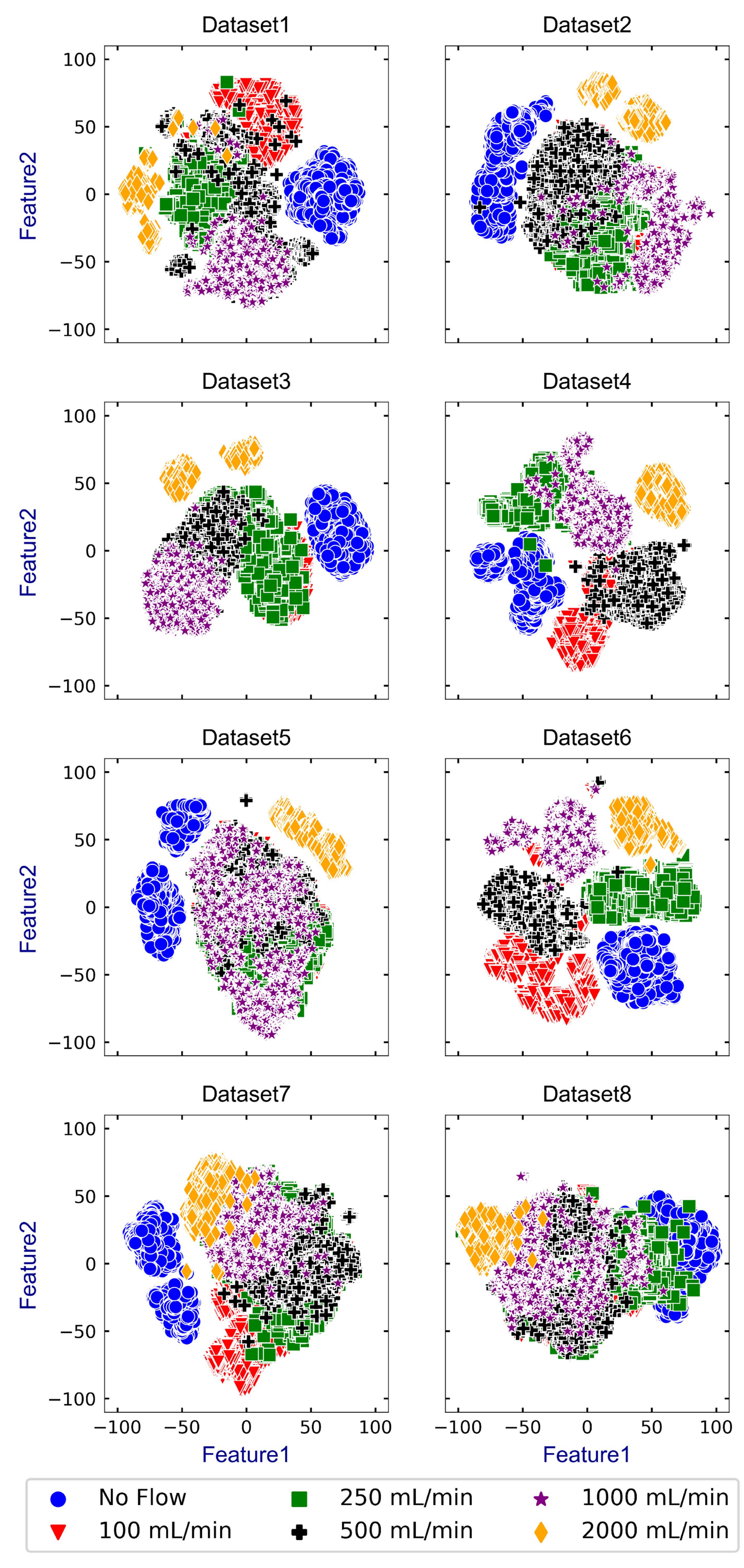}}
\caption{Visualization of each dataset in two dimensions using the t-SNE algorithm.}
\label{fig:fig5}
\vspace{-1.5em}
\end{figure}

Fig.~\ref{fig:fig5} shows various water flows in the collected datasets in two dimensions by t-SNE. The high flows of 2000 mL/min are well recognizable and distinguishable from the rest of the flows, including the zero flows shown in blue. On the other hand, the group of the other flows between 100 mL/min and 1000 mL/min overlap and cannot be distinguished from each other. However, even this group of flows can be clearly distinguished from zero flows. In other words, although different flows may not be distinguishable from each other, they are distinguishable from the zero flow cases, and the system can be used to detect water leakage over 100 mL/min. This is shown in Fig.~\ref{fig:fig5}, in which the zero flow, displayed in blue, can be easily distinguished from the other flows, displayed in red.
\vspace{0.2em}

\begin{figure}[!t]
\centerline{\includegraphics[width=\columnwidth]{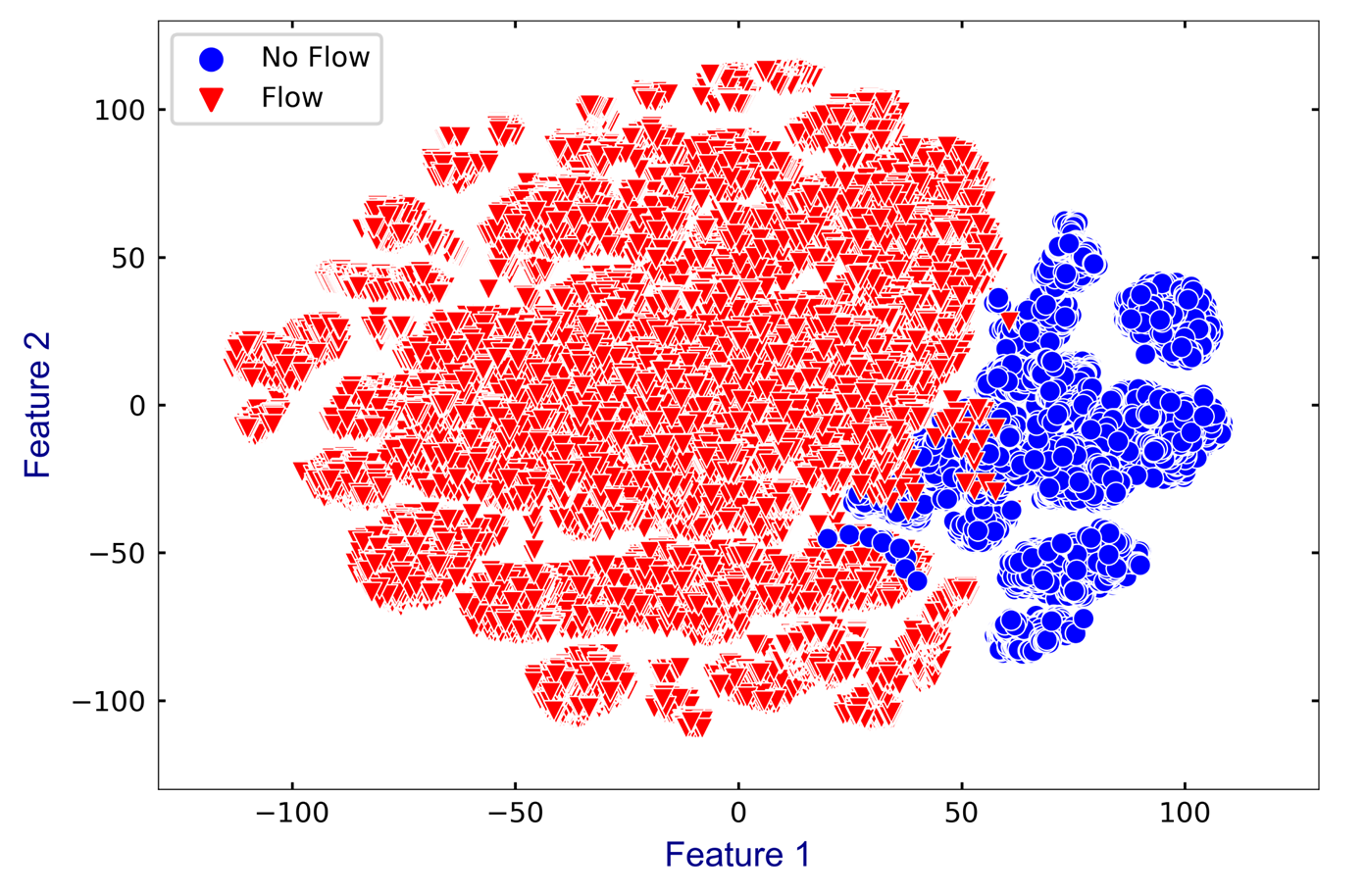}}
\caption{The status of zero flow (the blue one) versus other flows (the red one) in the collected datasets.}
\label{fig:fig6}
\end{figure}

\begin{table}[h]
\scriptsize
\centering
\caption{Comparison of the Implemented Models Accuracy Based on F1-score.}
\label{tab:tab3}
\begin{tabular}{p{25pt}p{150pt}p{65pt}}
\hline
\multicolumn{1}{c}{\textbf{Number}}          & \multicolumn{1}{c}{\textbf{Model Implemented}}     & \multicolumn{1}{c}{\textbf{F1-Score (Test)}} \\ \hline
\multicolumn{1}{c}{1}         & \multicolumn{1}{c}{Random Forest~\cite{b20}}          & \multicolumn{1}{c}{92.1\%}                \\
\multicolumn{1}{c}{2}     & \multicolumn{1}{c}{SVM}          & \multicolumn{1}{c}{94.9\%}                \\
\multicolumn{1}{c}{3} & \multicolumn{1}{c}{A Fully Connected Model}          & \multicolumn{1}{c}{95.1\%}                \\
\multicolumn{1}{c}{4} & \multicolumn{1}{c}{A Convolutional Model}          & \multicolumn{1}{c}{94.2\%}                \\
\multicolumn{1}{c}{\textbf{5}}                              & \multicolumn{1}{c}{\textbf{G-RBF Model with Fine-tuned parameters}} & \multicolumn{1}{c}{\textbf{97.7\%}}       \\
\multicolumn{1}{c}{6} & \multicolumn{1}{c}{A Fully Connected Model with OSM Loss~\cite{b22}}          & \multicolumn{1}{c}{97.3\%}  
\\ \hline
\end{tabular}
\end{table}

\begin{figure}[h]
\centerline{\includegraphics[width=\columnwidth]{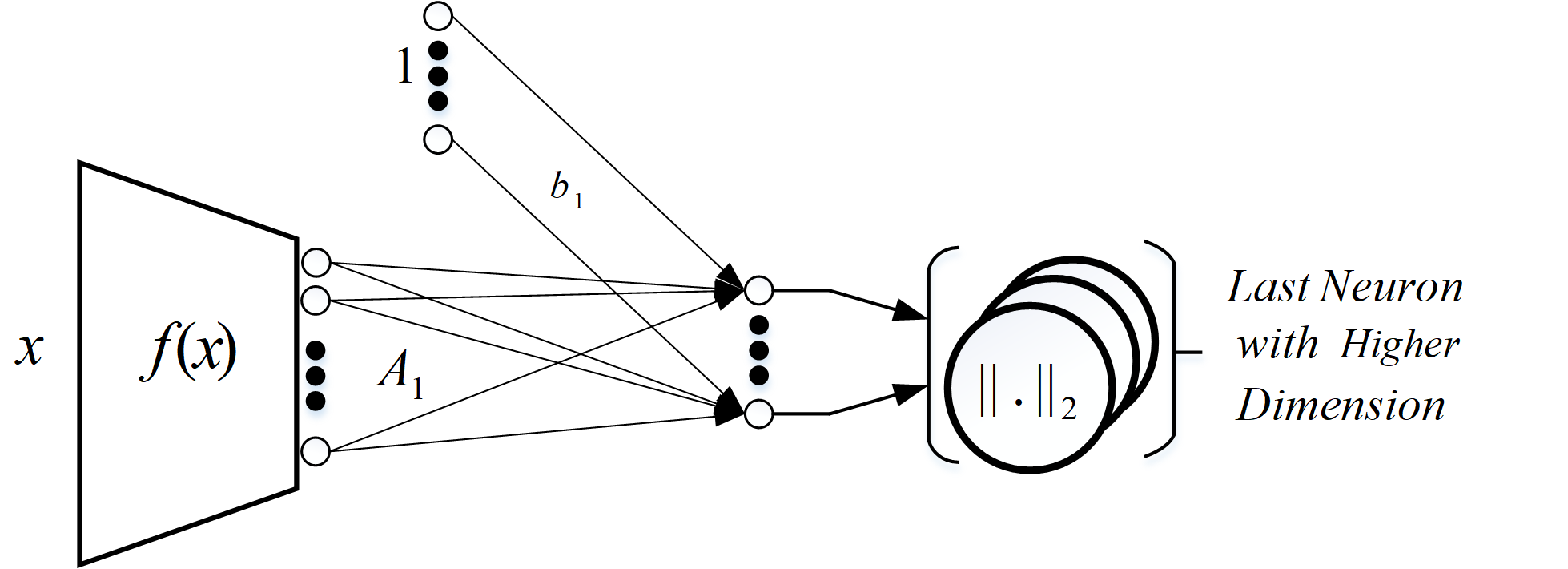}}
\caption{The structure of a deep G-RBF network with one class in 3 dimensions in which one RBF is implemented in its output layer to classify no-flow and flow classes.}
\label{fig:fig7}
\end{figure}

We implemented different algorithms such as Random Forst, SVM, and G-RBF, reported in Table.~\ref{tab:tab3}, to find the best classification approach to distinguish between water leakage cases from non-water leakage cases. The results show that the Deep Gaussian RBF (G-RBF) model \cite{b21} outperformed the other algorithm. In Fig.~\ref{fig:fig7}, the structure of the G-RBF model is shown in which features pass through an arbitrary model f(x). Then, in the final layer, the Radial Basis Function (RBF) is implemented to detect no flow samples in front of rest samples as flow samples; this model can separate zero flows from whatever is called non-zero flow with its rejection property and put them in the non-zero or rejection class.

The most important part of this network is its last layer, which is responsible for implementing the RBF function. The objective function is shown in Eq.~\ref{eq:eq1} from \cite{b21}, in which $y_i$ corresponds to the actual class of input $x^{(i)}$, and $d_j$ represents the RBF distance of the samples. In addition, $\lambda$ is the hyperparameter that should be adjusted according to the data type, representing the distance or boundary between classes.

\begin{equation}
J_{ml} = \sum_{i=1}^{N}(d_{y_i}(x^{(i)}) + \sum_{j \notin y_i}max(0, \lambda - d_j(x^{(i)}))).
\label{eq:eq1}
\end{equation}

The objective function, shown in Eq.~\ref{eq:eq1}, intends to bring samples related to zero flow closer to the center and move samples related to non-zero flow away from the center based on $\lambda$. Eq.~\ref{eq:eq1} illustrates the simple form of the objective function, but in order to implement it practically, Eq.~\ref{eq:eq1} has been used in which $Algorithmic$ functions replace $Max$ functions.

\vspace{0pt}
\begin{equation}
\resizebox{.9\columnwidth}{!}{$J_{SoftML} = \sum_{i=1}^{N}(d_{y_i}(x^{(i)}) + \sum_{j \notin y_i}log(1 + exp(\lambda - d_j(x^{(i)})))).$}
\label{eq:eq2}
\end{equation}
\vspace{1.0em}

The result of different classifiers, presented in the next section (Table~\ref{tab:tab3}), shows that this model has excellent accuracy compared to others.

\vspace{0.5em}

\subsection{Results}

\begin{figure}[!t]
\centerline{\includegraphics[width=\columnwidth]{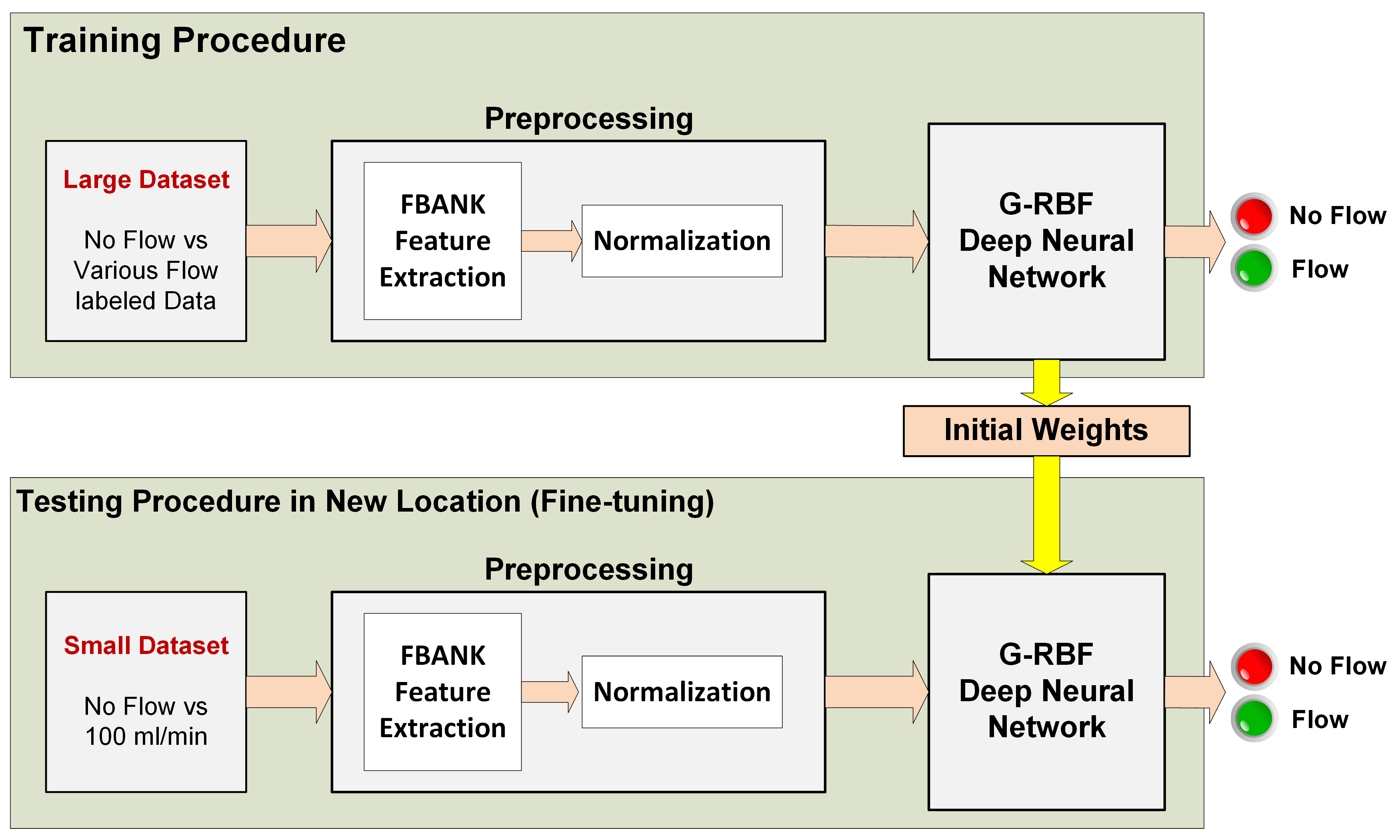}}
\caption{Training and testing process block diagram used in the proposed water flow detection system.}
\label{fig:fig8}
\end{figure}

Fig.~\ref{fig:fig8} displays the block diagram of the training and calibration/fine-tuning phases of the proposed system. In the above row, the training phase of the proposed model is illustrated. Since the proposed device must be used in different places, its performance to detect water flow will change because of differences in water supply structures. In the calibration/fine-tuning phase, a small dataset from zero and 100 mL/min water flow data would be collected to fine-tune parameters, as shown in the bottom row of Fig.~\ref{fig:fig8}.

In Table. \ref{tab:tab3}, the results of different models are shown. From a total of nine datasets that have been collected, eight datasets have been used to train models and set hyperparameters by using the cross-validation method. Since models have not seen dataset number nine, it has been used to test models. Moreover, all datasets are asymmetrical, and samples with non-zero flow labels are more than samples with zero flow labels, so the F1-score method has been used to report the accuracy.

To test the reliability, it has been installed and tested in three different locations. Table.~\ref{tab:tab4} presents the result of the proposed method, using the G-RBF approach, in the three locations after the fine-tuning phase.

\vspace{0.5em}
\begin{table}[h]
\scriptsize
\centering
\caption{The Accuracy of the Proposed System in New Locations After Fine-tuning Model Parameters.}
\label{tab:tab4}
\resizebox{250pt}{!}{
\begin{tabular}{ccc}
\hline
\textbf{Location}          & \textbf{Description}     & \textbf{F1-Score Result} \\ \hline
1         & Fine-tuned Trained Model in Location 1          & 91.5\%                \\
2     & Fine-tuned Trained Model in Location 2          & 91.4\%                \\
3 & Fine-tuned Trained Model in Location 3          & 94.5\%             
\\ \hline
\end{tabular}}
\end{table}
\vspace{0.5em}

\section{Conclusion}
In this paper, we proposed a low-cost and non-invasive system to detect leakage. The system relies on the fact that there should be zero water flow throughout a certain period of time. Thus, if there is a water leakage over a certain amount of water, then it can be detected by the proposed system.

The novel idea behind the proposed system is using mechanical sound signal amplifiers to gather sound signals generated by small water flows. Furthermore, the system is designed in a way to easily and effectively transfer sound signals from water pipes to the microphone. Finally, modern machine learning methods are employed to distinguish between water leakage over 100 mL/min and no water leakage.
\vspace{0.5em}

\scriptsize
\bibliographystyle{IEEEtran}
\bibliography{Main}

\vskip 0pt plus -1fil

\begin{IEEEbiography}[{\includegraphics[width=1in,height=1.25in,clip,keepaspectratio]{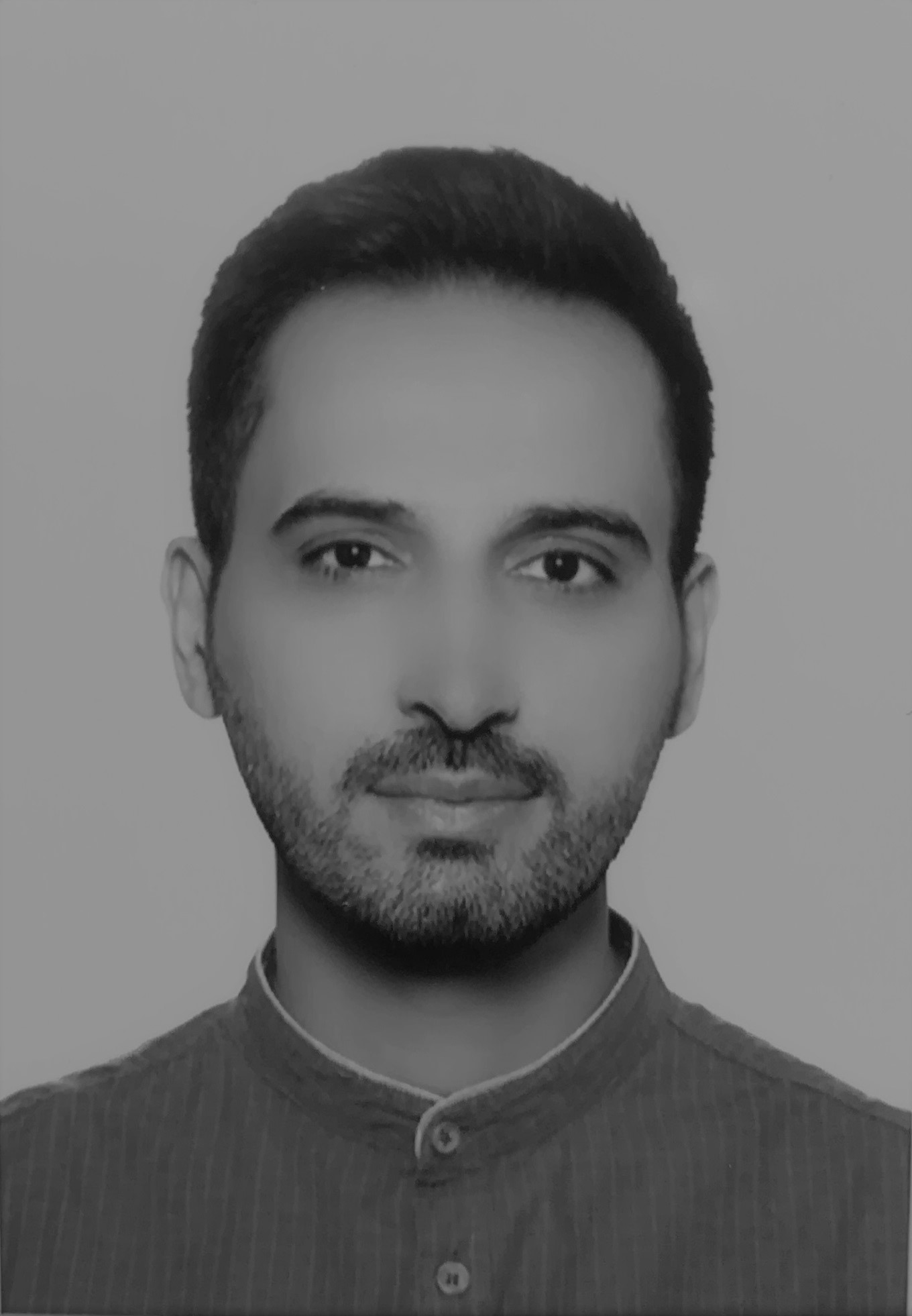}}]{Hossein Pourmehrani} received his B.Sc. degree in hardware engineering from the University of Tehran, Iran, in 2011 and his M.Sc. degree in Artificial Intelligence and Robotics from the University of Tehran, Iran, in 2022. He is currently a Ph.D. student at the University of Maryland Baltimore County (UMBC) and a member of the Secure, Reliable, and Trusted Systems (SECRETS) research lab since joining UMBC in 2023, USA. His current research focuses on hardware security, particularly Side-channel analysis, and fault injection attacks and their countermeasures on Machine learning hardware as well as designing efficient and secure hardware accelerators for machine learning models.
\end{IEEEbiography}

\vskip -1.5\baselineskip plus -1fil

\begin{IEEEbiography}[{\includegraphics[width=1in,height=1.25in,clip,keepaspectratio]{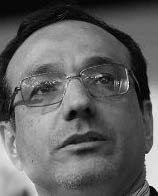}}]{Hadi Moradi}
received his B.Sc. degree in electrical and electronics engineering in 1989 from the University of Tehran (UT). He received his M.Sc. and Ph.D. degrees in robotics and intelligent systems from the University of Southern California (USC) in 1994 and 1999, respectively. He has experience in industry and education since 1997, working at FEM Engineering for two years and at USC and SKKU in South Korea between 2000 and 2008. In 2008 he joined the School of Electrical and Computer Engineering at UT. His research interest includes computational psychology using games, intelligent systems, and robots. He has been working with children with autism since 2010. Dr. Moradi received several awards, including the best paper award at CLAWAR2008 and a finalist for the best paper award in ROBIO 2011. He also received several awards for his games and cognitive screening and rehabilitation methods. He has been heavily involved with IEEE activities since 2004 in positions such as the chair of the IEEE Iran section and the co-chair of the service robots technical committee of the Robotics and Automation Society.
\end{IEEEbiography}

\vskip -1.5\baselineskip plus -1fil

\begin{IEEEbiography}[{\includegraphics[width=1in,height=1.25in,clip,keepaspectratio]{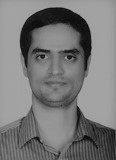}}]{Reshad Hosseini} received his B.Sc. degree in electrical engineering (telecommunication) from the University of Tehran, Tehran, Iran, in 2004, his M.Sc. degree in biomedical engineering (bioelectric) from the Amirkabir University of Technology, Tehran, in 2007, and his Ph.D. degree from the Faculty of Electrical Engineering and Computer Science, Technical University of Berlin, Berlin, Germany, in 2012. He did his Ph.D. research at the Max Planck Institute for Biological Cybernetics, Tuebingen, Germany. He is currently an Assistant Professor at the School of Electrical and Computer Engineering, College of Engineering, University of Tehran. His professional interests are machine learning, signal processing, and computational vision. He is particularly interested in the mathematical foundation of these fields, such as differential geometry, optimization, functional analysis, and statistics. His current research interests include representation learning using deep models, visual recognition using deep learning, sequence-to-sequence models, generative models, manifold optimization, large-scale mixture models, 3-D reconstruction, and accelerating reinforcement learning.
\end{IEEEbiography}

\end{document}